# Manufacturing of A micro probe using supersonic aided electrolysis process

R.F. Shyu, Feng-Tsai Weng, and Chi-Ting Ho
Institute of Mechanical and Electrical-mechanical Engineering
National Formosa University, Taiwan
886-5-6315398
ftweng@nfu.edu.tw

*Abstract*-In this paper, a practical micromachining technology was applied for the fabrication of a micro probe using a complex nontraditional machining process. A series process was combined to machine tungsten carbide rods from original dimension. The original dimension of tungsten carbide rods was 3mm; the rods were ground to a fixed-dimension of 50μm using precision grinding machine in first step. And then, the rod could be machined to a middle-dimension of 20μm by electrolysis. A final desired micro dimension can be achieved using supersonic aided electrolysis. High-aspect-ratio of micro tungsten carbide rod was easily obtained by this process. Surface roughness of the sample with supersonic aided agitation was compared with that with no agitation in electrolysis. The machined surface of the sample is very smooth due to ionized particles of anode could be removed by supersonic aided agitation during electrolysis. Deep micro holes can also be achieved by the machined high-aspect-ratio tungsten carbide rod using EDM process. A micro probe of a ball shape at the end was processed by proposed supersonic aided electrolysis machining process.

Keywords: Micromachining, Micro probe, Electrode, Electrolysis, Supersonic.

## I. INTRODUCTION

Micro-electrical-mechanical-system (MEMS) technology becomes important due to the growing electro-optical industry. The cost amount of products is very significant in the industry fields of semi-conductor, precision fabrication, optical, and biotechnology in recent years. Electrical discharge machining (EDM) process which can be used for micromachining usually uses graphite or a metal tool to generate high frequency sparks between electrode and the work piece in a dielectric fluid. A rapid methodology has been proposed to form 3D electrode using the electroplating of EDM [1]. However the machining time of the process is too long. Since wire electrical discharge grinding (WEDG) system was developed to process microelectrode [2], the potential micromachining in EDM process is noticed from then on. Researchers used the same idea of WEDG to machine a 4.3μm microelectrode with a aspect ratio of 10 [3]. The electrode can be used to drill micro hole in EDM process, and to machine the silicon wafer. However, the wear of tungsten carbide is large, the aspect-ratio is still low, and the bending of micro electrode can occasionally happen due to the machining stress. Copper can be etched by $FeCl_3$ or $CuCl_2$ enchant [4-6], and the thickness of the TiNi foil could be etched to about 10μm by a proper volume ratio of $HF/HNO_3/H_2O$ etching solution [7]. Although chemical machining (CHM) can be used to machine a metal rod to a small size, but the surface of the work piece is uneven [8]. Electrochemical machining (ECM) is a well known for the non-traditional machining process. Electroplating, electro grinding, and electrolysis are all part of applications of ECM. Nano electrode of the tungsten with a tip of about 500–800 nm could be achieved by the anodic etching [9-10]. Although the result of experiment shows this process has capability to produce the nanometer scale electrode, but the aspect-ratio of the electrode is still not large enough, and their wire materials are too soft that can not be widely adapted in the industry. Applications of Supersonic aided machining combined with other machining processes, such as turning, milling [12], EDM [13], ECM, CHM, grinding, and polishing, can provides better performance, where the removal rate is higher, the cutting force and the tool wear are reduced [11], the tool life and the machined surface roughness are improved, better fluid diffusion and propagation can be obtained by the disturbance of the supersonic [14].

Tungsten carbide has the good stiffness and the heat resistance, and it can be machined to less than 10μm by precision grinding process. It can provide excellent electrode for the micro-EDM, but the aspect ratio is still not large enough. In this study, a composite processing technology includes precision grinding, super sonic agitation, and electrolysis were combined to process the tungsten carbide material. Tungsten carbide alloy rods of diameter 3.0 mm could be machined to the desired micro size by this process.

## II. EXPERIMENTAL EQUIPMENTS AND PROCEDURE

The electrolysis system for machining micro electrodes contained a D.C. power supply and a container for electrolyte. The power supply used in the experiment is a model CT-3032D, which has an output voltage of 0-30 volt, and an output current range of 20mA –2A. It is made in Taiwan by the Sampo Corporation. A supersonic cleaner was used for cleaning the work piece during electrolysis. This machine offers an output of 50-60Hz, at max 235W power, and an output of 100W, at an operating frequency of 42 KHz. This machine is manufactured by Branson sonic Corporation, USA. (Model 2501R-DTH). The vibration amplitude of this machine is about 30μm. A precision grinding machine was used for the machining in first step. The grinding machine is manufactured in Germany. It provides highly automatic NC machining capability in micro machining.

The experimental procedures include: 1) 3mm tungsten carbide electrodes were processed by the precision grinding to two different sizes, one of a ball of diameter 150μm at the end, and another of rod 50μm at the end. 2) Machined microelectrodes were diverged separately into the electrolyte, and were connected as an anode; pure copper was connected as a cathode for electrolysis. The electrolyte was copper sulfate hydrate solution. The concentrate of electrolyte for electrolysis in experiments is $CuSO_4 \cdot 5H_2O$ (50 g/l), $CuCl_2 \cdot 2H_2O$ (0.4 g/l), $H_2SO_4$ (15 g/l). Initial. In the machining of electrolysis, the anode electrode is eroded in accordance with the Faradays' law. Reaction of anode is metal ionized in the electrolyte from electrode, and cathode is iron deposited to the electrode.

## III. RESULTS AND DISCUSSION

Table1 shows the comparison of the surface roughness of the samples that were processed by electrolysis with supersonic aided agitation and that with no agitation. Initial voltage was 10V, Initial current was 15mA. Tungsten carbide electrodes were machined from 50μm to 20μm. From results, the surface roughness with no agitation is much higher than that with agitation, and the machined time with agitation is also quicker than that with no agitation. The reason may be that electrolysis generated particles on anode can be automatically separated from the electrode with agitation of supersonic. From Fig.1 to Fig.4, SEM photographs show different machined diameters with different machined voltage, current, and time. Tungsten carbide microelectrode was first machined from 3mm to 50μm by precision grinding (Fig.1). Electrode length is about 150μm. Fig.2 and Fig.4 shows the different diameter of the tungsten carbide microelectrode from 20μm to 2μm, and they show good quality machining results by supersonic agitation. Fig.2 and Fig.3 shows the SEM photograph of a micro electrode with no agitation during machining, the particles around the electrode can be effectively removed by supersonic agitation. High aspect ratio of the micro tungsten carbide rod can be obtained by this process. Fig.5 shows the SEM photograph of a micro hole of a copper plate machined by a fine electrode of 10μm by EDM machining. The thickness of the copper plate is 50μm. The current is 0.3A, the voltage is 35V, and the machining time is about 1min. Fig.6 shows SEM photo of a micro probe by supersonic aided electrolysis. The machined current is 0.2mA, Voltage is 10V. The machine time is about 20min. Fig.7 shows SEM photo of a micro probe by supersonic aided electrolysis. Dimension of micro probe with a tip of about 20nm. The machined current is 0.2mA, Voltage is 10V. The machine time is about 20min. The diameter before machining was 20μm. Fig.8 shows SEM photo of Deep micro hole machined by a 20μm tungsten carbide electrode using EDM process. Material is a 0.2mm stainless plate. The machined current is 0.3A, Voltage is 35V. The machine time is about 5min.

## IV. CONCLUSIONS

Micro electrodes can be processed to a desired micron size by varies machining time in a fixed small voltage and current. Electrodes can be processed to a very small size with no bending. This technology has potential in the nano scale electrode machining. Deep micro holes can also be achieved by the machined high-aspect-ratio tungsten carbide rod using EDM process. A micro probe of ball diameter was processed to be 80μm at the end by proposed supersonic aided electrolysis machining process. The micro probe can be used as a sensor for measurement equipments. Dimension of micro probe with a tip of about 20nm could be obtained at the tip of the end by this process.

Table 1
Comparison of copper surface roughness which the samples were processed by electrolysis and electroplating of supersonic aided agitation and no agitation. The voltage was 10V, and the current was 15mA.

|  | no agitation | supersonic aided agitation |
|---|---|---|
| Surface roughness of electrolysis | 2.51μm Ra | 1.67μm Ra |
| Machining time | 140min | 92min |

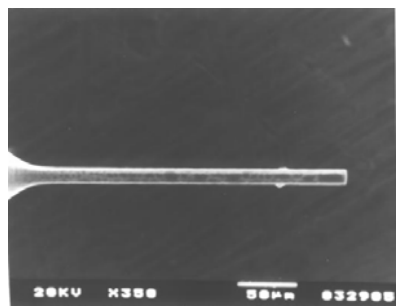

Fig.1 SEM photo of tungsten carbide microelectrode machined by precision grinding. The diameter is 50μm. The length is 150μm.

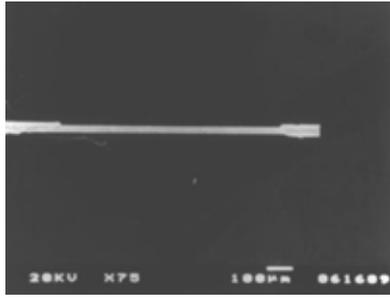

Fig.2 SEM photo of 20µm tungsten carbide microelectrode. This electrode is obtained by 0.5mA and 10V for 1/2hr. The original diameter before machining was 50µm.

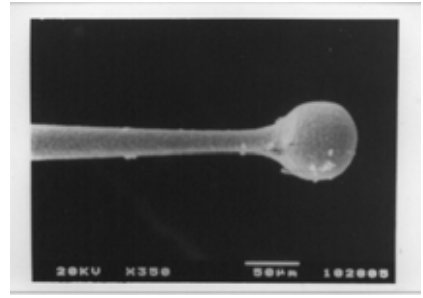

Fig.6 SEM photo of a micro probe by electrolysis. The machined current is 0.2mA, Voltage is 10V. The machine time is about 20min.

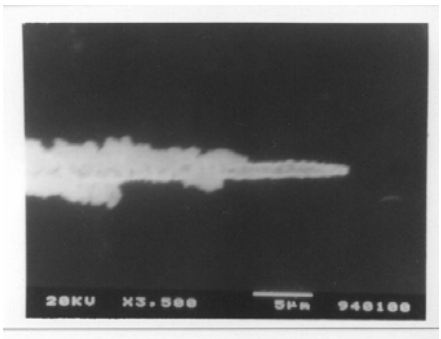

Fig.3 SEM photo of 10µm tungsten carbide. This electrode is obtained by 0.2mA and 10V for 45min. The original diameter before machining was 20µm.

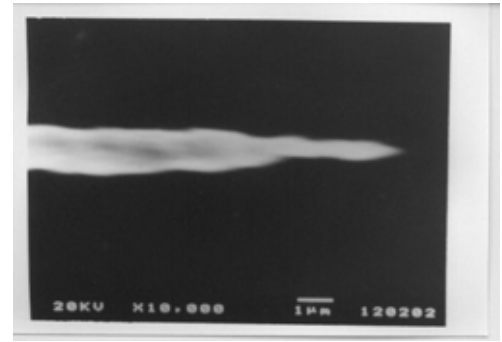

Fig.7 SEM photo of a micro probe by electrolysis. Dimension of micro probe with a tip of about 20nm. The machined current is 0.2mA, Voltage is 10V. The machine time is about 20min. The diameter before machining was 20µm.

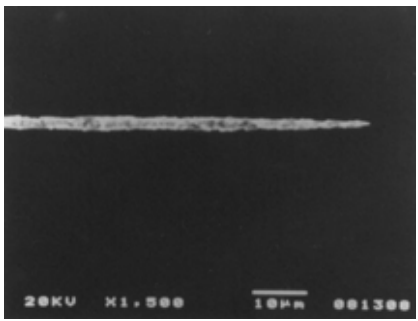

Fig.4 SEM photo of 2µm tungsten carbide. This electrode is obtained by 0.1mA and 10V for 1hr. The original diameter before machining was 20µm.

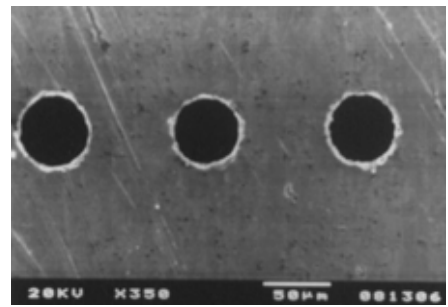

Fig.8 SEM photo of Deep micro hole machined by a 20µm tungsten carbide electrode using EDM process. Material is a 0.2mm stainless plate. The machined current is 0.3A, Voltage is 35V. The machine time is about 5min.

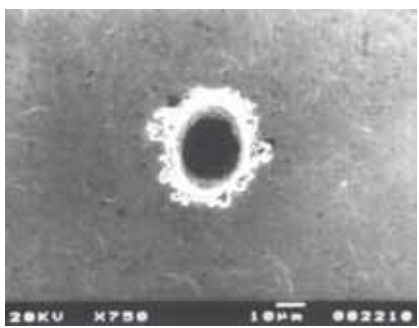

Fig.5 SEM photo of a micro hole machined by a fine electrode of 10µm by EDM machining. The copper plate is 50µm.The current is 0.3A, Voltage is 35V. The maching time is about 1min.

ACKNOWLEDGMENT

The authors are grateful to the National Science Council of Republic of China for the support of this research grant No 93-2622-E-150-051-CC3.

REFERENCES

[1] D. E. Dimla, N. Hopkinson, H. Rothe., "Investigation of complex rapid EDM electrodes for rapid tooling applications", The international journal of advanced manufacturing technology. No. 973, Vol. 23. 2004, pp. 249-255.

[2] T. Masuzawa, M. Fujino, K. Kobayashi, T. Suzuki, Wire Electro-Discharge Grinding for micro machining. Ann CIRP 34(1), pp. 431-434. 1985.

[3] T. Masaki, K. Kawata and T. Masuzawa, "Micro electro-discharge machining and its application", Micro Electro Mechanical Systems, Proceedings, an Investigation of Micro Structures, Sensors, Actuators, Machines and Robots. IEEE, pp.21 –26. 1990.

[4] Liang-Tau Chang. "Studies on Cupric Chloride Etching Solution",Journal of Technology, Vol 9, No. 3, pp291-297, 1994.

[5] Madhav Datta, Derek Harris, "Electrochemical Micromachining: An Environmentally Friendly, High Speed Processing Technology", Electrochimica, Vol. 42, Nos 20-22, pp. 3007-3013. 1997.

[6] Cai Jian,Ma Jusheng,Wang Gangqiang,Tang Xiangyun, "Effects on etching rates of copper in ferric chloride solutions", IEMT/IMC Symposium, 15-17 April 1998 pp.144-148. 1998.

[7] Z. Z. Chen, S. K. Wu, "Chemicl Machined Thin Foils of TiNi Shape Memory Alloy", Materials Chemistry and Physics, 58, pp. 162-165. 1999.

[8] Feng-Tsai Weng, "Fabrication of microelectrodes for EDM machining by a combined etching process", Journal of Micromechanics and Microengineering. volume 14, issue 5, N1 - N4. 2004.

[9] B. Bhattacharyya, B. Doloi, P.s. Sridhar., "Electrochemical micro-machining: new possibilities for micro-manufacturing", Journal of Materials processing Technology. 113, pp301-305. 2001.

[10] Katsunobu Yamamoto, Guoyue Shi , Tianshu Zhou, Fan Xu, Min Zhu, Min Liu, Takeshi Kato, Ji-Ye Jin, and Litong Jin. "Solid-state pH ultramicrosensor based on a tungstic oxide film fabricated on a tungsten nanoelectrode and its application to the study of endothelial cells", Analytica Chimica Acta 480 109–117. 2003.

[11] Q. H. Zhang, J. H. Zhang, Z. X. Jia, J. L. Sun., "Material Removal-Rate analysis in the ultrasonic machining of engineering ceramics". Journal of Materials processing Technology. 88 . pp180-184. 1999.

[12] Z. J. Pei, P. M. Ferreira, S. G. Kapoor, M. Hasekorn., "Rotary ultrasonic machining for face milling of ceramics". Int. J. Tools Manufact. Vol. 35. No. 7. Pp.1033-1046. 1995.

[13] Z. N. Guo, T. C. Lee, T. M. Yue, W. S. Lau, "A Study of Ultrasonic-aided Wire Electrical Discharge Machining", Journal of Materials processing Technology, 63, pp. 823-828. 1997.

[14] W. Roetzel, B. Spang, X. Luo, S.K. Das., "Propagation of the third sound wave in fluid: hypothesis and theoretical foundation", International Journal of Heat and Mass Transfer 41 pp2769-2780. 1998.